\documentclass[conference]{IEEEtran}
\IEEEoverridecommandlockouts
\usepackage{cite}
\usepackage{amsmath,amssymb,amsfonts}
\usepackage{algorithmic}
\usepackage{graphicx}
\usepackage{textcomp}
\usepackage{xcolor}
\usepackage{url}
\usepackage{booktabs} 
\usepackage{hyperref}
\usepackage{multirow}
\usepackage{quoting}

\def\BibTeX{{\rm B\kern-.05em{\sc i\kern-.025em b}\kern-.08em
    T\kern-.1667em\lower.7ex\hbox{E}\kern-.125emX}}
\begin{document}

\title{Enhancing Lie Detection Accuracy: A Comparative Study of Classic ML, CNN, and GCN Models using Audio-Visual Features}

\author{
    \IEEEauthorblockA{\textit{Abdelrahman Abdelwahab} \\
    abdelrahman.a.abdelaziz28@gmail.com \\ STEM high school for boys 6-October}
    \and
    \IEEEauthorblockA{\textit{Akshaj Vishnubhatla} \\
   akshajvishnubhatla@gmail.com \\ Briar Woods HS}
    \and
    \IEEEauthorblockA{\textit{Ayaan Vaswani} \\
    ayaanvaswani@gmail.com \\Bellarmine College Preparatory}
    \and

    \IEEEauthorblockA{\textit{Advait Bharathulwar} \\
   advaitbharathulwar@gmail.com \\ John P Stevens High School}
    \and
    \IEEEauthorblockA{\textit{Arnav Kommaraju} \\ arnavkommaraju@gmail.com \\Edison High School}

    \and
    \IEEEauthorblockA{\textit{Bohan Yu} \\
   ybhtim@berkeley.edu \\ UC Berkley}

}

\maketitle

\begin{abstract}
Inaccuracies in polygraph tests often lead to wrongful convictions, false information, and bias, which have significant consequences for both legal and political systems. Recently, analyzing facial micro-expressions has emerged as a method to detect deception; however, current models have not reached high accuracy and generalizability. The purpose of this paper is to aid in remedying these problems. The unique multimodal transformer architecture used in this paper improves upon previous approaches by using auditory input, visual facial micro-expressions, and manually transcribed gesture annotations, moving closer to a reliable non-invasive lie detection model. Visual and auditory features were extracted using Vision Transformer and OpenSmile models respectively, which were then concatenated with the transcriptions of participants’ micro-expressions and gestures. Various models were trained for classification instances of lies and truth using these processed and concatenated features. The CNN Conv1D multimodal model achieved a 95.4\% average accuracy. However, further research is still required to create higher-quality datasets and even more generalized models for more diverse applications. 
\end{abstract}


\section{Introduction}
Lie detection has been a recurring focus of research and technological innovation in law enforcement and criminal justice. According to a survey conducted by the University of Wisconsin-La Crosse, about 75\% of survey respondents reported telling zero to two lies per day; lying comprised 7\% of total communication, with 79\% of the lies being told face-to-face and 21\% being mediated \cite {morgan2021}.

Current technologies, such as polygraphs, have focused on biological responses like blood pressure to detect lies. However, these methods are unpredictable and easily flawed. Recently, research has begun to focus on various other indicators of deception, including facial micro-expressions and audio cues \cite{mahon2007}. Facial micro-expressions (ME) are intentional or involuntary localized and momentary movements of the face, usually lasting less than 500 milliseconds \cite{merghani2018}.

Despite advancements in lie detection techniques, traditional methods remain intrusive, subjective, and often inaccurate. Detecting deception through ME and speech analysis presents a significant challenge due to the subtle and brief nature of these cues. As shown in Table \ref{tab:lie_detection_accuracy}, traditional methods have high variance and relatively low accuracy. This study aims to address these limitations by developing a non-intrusive, objective, and highly accurate method for detecting deception using both ME and audio signals. Accurate lie detection is crucial in various fields, including security, legal systems, and psychological evaluations.
The primary objective of this study is to establish an AI model that can differentiate between truth and deception with high accuracy by analyzing audio, visual cues in videos, and extracted gestures. Audio dialogue, visuals, and gestures all help to distinguish between deception and truthfulness, making them important features to consider \cite{Galinsky2022}. Therefore, the Real-life Deception Detection Dataset from the University of Michigan was used, which includes 121 videos of deception and truthfulness and a CSV file for gestures. Visuals and audio were extracted from the videos, and OpenSMILE and Vision Transformer (ViT) were used to extract features from audio and video, respectively. Classical machine learning models like Random Forest Classifiers and Logistic Regression can serve as accurate baseline references for a binary classification task like truth and lie. Yet to build off of that, by leveraging advanced neural network models, such as Conv1D, Graph Convolutional Networks (GCN), and CNN LSTM, the accuracy can be increased.

\begin{table}[h]
    \caption{Estimated accuracy of different test types in detecting deception and truthfulness}
    \label{tab:lie_detection_accuracy}
    \centering
    \begin{tabular}{|l|c|c|}
        \hline
        \textbf{Test type} & \textbf{Detecting deception} & \textbf{Detecting truthfulness} \\
        \hline
        \multicolumn{3}{|l|}{\textbf{Laboratory studies}} \\
        \hline
        CQT -- Polygraph & 74\%--82\% & 60\%--83\% \\
        CIT -- Polygraph & 76\%--88\% & 83\%--97\% \\
        ERP & 68\% & 82\% \\
        fMRI & 84\% & 81\% \\
        \hline
        \multicolumn{3}{|l|}{\textbf{Field studies}} \\
        \hline
        CQT -- Polygraph & 84\%--89\% & 59\%--75\% \\
        CIT -- Polygraph & 42\%--76\% & 94\%--98\% \\
        \hline
    \end{tabular}

\end{table}

This study addresses the following research questions: How effective is the proposed AI model in detecting lies compared to traditional methods and some recent AI models? Which features carry the highest weights in prediction?
Deception detection technology has the potential to revolutionize various fields. In law enforcement, it could improve interrogation outcomes and border security by identifying deceptive behavior. In the legal system, it could be utilized to assess the credibility of courtroom testimonies and negotiations. Additionally, applying this technology to financial services could aid in detecting fraudulent claims and reducing the risk of financial fraud.

Previous studies have experimented with various machine learning models. For instance, a study by Soldner et al. implemented the Random Forest model, achieving the best accuracy of 69\%, as shown in \ref{table:2} \cite{soldner2019}. Insights from this paper suggest expanding our dataset and exploring additional modalities to enhance the model's accuracy and reliability in lie detection. Furthermore, Random Forest, being a machine learning technique, cannot handle complex relations as well as multimodal data, which is a limitation of the mentioned study. Moreover, most traditional AI models fall short in reliability and accuracy, often leading to false positives or negatives \cite {li2022}. A study conducted by the University of Michigan in 2015 analyzed trial videos using micro-facial expressions and achieved a rudimentary accuracy rate of 83.05\% using neural networks [6]. Aligning different data types and achieving 83.05\% accuracy are two main advantages of the study.

\begin{table}[h!]
\centering
\caption{ Best Results Of Study [2].}
\label{table:2}
\resizebox{0.2\textwidth}{!}{%
\begin{tabular}{lcc}
\toprule
Features & Acc. \\
\midrule
Linguistic & 66\% \\
Dialog & 57\% \\
Non-verbal & 61\% \\
All Features & 69\% \\
\bottomrule
\end{tabular}}
\end{table}

This paper is organized as follows: analyzing previous work, discussing the paper’s methods (data collection, data analysis, feature extraction, and implementation guide for the tested models), presenting the results of different tested models, comparing the paper’s results with other studies using the same dataset, and providing a discussion including limitations and recommendations. The paper concludes with a summary of key findings and a look forward.

\section{Literature Review}
\subsection{Prior solutions}
The paper, titled \textit{Facial Micro-Expression Recognition Based on Deep Local-Holistic Network}, introduces a Deep Local-Holistic Network (DLHN) for micro-expression recognition, comprising two sub-networks: the Hierarchical Convolutional Recurrent Neural Network (HCRNN) and the Robust Principal Component Analysis Recurrent Neural Network (RPRNN). HCRNN captures local spatiotemporal features using CNNs and BRNNs, whereas RPRNN extracts global sparse features using RPCA and BLSTM networks. The DLHN was evaluated on four combined datasets (CASME I, CASME II, CAS(ME)2, and SAMM), achieving an accuracy of 60.31\%, outperforming several state-of-the-art methods \cite{li2022}.

In the study by Feng (2021), titled \textit{DeepLie: Detect Lies with Facial Expression (Computer Vision)}, the author developed a deep learning-based approach to lie detection using facial micro-expressions in video streams. The method employs a Siamese network architecture with triplet loss to effectively distinguish between truthful and deceptive expressions. Key components include the use of CNNs for feature extraction and a GRU-based RNN for sequence learning. The model achieved an 81.82\% accuracy on the validation dataset. However, the study highlighted limitations due to the small size of the dataset, which may hinder the model's generalizability. The author suggests that future work should focus on incorporating multi-modal data (e.g., audio and text) and expanding the dataset to include more diverse scenarios \cite{feng2021deeplie}.

The \textit{Hybrid Machine Learning Model for Lie Detection} research dataset included thermal images. The study employed a hybrid machine-learning approach that combined the strengths of CNNs and SVMs. CNNs were used to extract relevant features from the input data automatically. These extracted features were then fed into an SVM for classification. As a result, the hybrid model demonstrated an accuracy of approximately 58\%. However, the complexity of the hybrid model, combining CNNs and SVMs, could lead to higher computational costs and increased difficulty in scaling the approach to larger datasets \cite{Dhabarde}.

The \textit{Audio-Visual Deception Detection Using the DOLOS Dataset} study used the DOLOS dataset that combined synchronized audio and video signals. The study employed ImageNet pre-trained ViT as the backbone network for the visual modality and tokenized face images with a 2D-CNN module, resulting in a feature with a dimension of 64 × 256. For the audio modality, the study adopted the pre-trained W2V2 model. The raw audio was tokenized by the 1D-CNN module, resulting in a feature size of 64 × 512 for each audio sample. The Plug-in Audio-Visual Fusion model and multi-task learning achieved an accuracy of 66.84\% \cite{guo2023audiovisual}.

Numerous studies have also been conducted using the Real-Life Trial Dataset. Camara et al(2024) summarizes many different studies conducted using this dataset. Ding et al. (2019) used a CNN model. ResNet served as the backbone of the face expression and computed the temporal feature maps. This was used to achieve this accuracy in conjunction with a Generative adversarial network (GAN). This model was reported to achieve a 97\% accuracy, which is the highest among current deep learning models \cite{Ding}. Among non-deep learning models, the use of SVM in Carissimi et al. (2018) gave an accuracy of 99\%. This model used features from AlexNet. In Wu et al (2018), a logistic regression model using Improved Dense Trajectory (IDT) features was reported to have an accuracy of 92.21\%. Other studies were also conducted on this dataset with high accuracy; however, their methodology was vague and could not be readily replicated \cite{Camara}. A key problem to note is that current studies have used varying features when working with different models, making comparison between models difficult. This study attempts to better display this comparison by using the same features across models.

\subsection{Models Overview}
\subsubsection{Logistic Regression}
Two classical machine learning models were considered for this study: Logistic Regression and Random Forest Classifier. Logistic regression is a binary classification technique based on the sigmoid function. This function is used to weight features in a way that returns a value from 0 through 1\cite{saidi2021}. 
    The logistic regression function can be defined with the equation \cite{Yang2019}
    \[
P(y=1 \mid X) = \sigma(z) = \frac{1}{1 + e^{-z}}
\]

\begin{figure}[h!]
\centering
\includegraphics[width=0.5\textwidth]{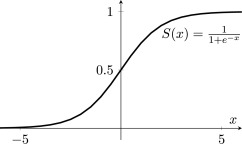}
\caption{Diagram of Sigmoid function \cite {Yang2019}}
\label{fig: sigmoidfunction}
\end{figure}

Where:
\begin{itemize}
    \item \( P(X) \) is the probability that the outcome \( y \) is 1 given the input \( X \).
    \item \( z \) is the linear combination of input features and their corresponding weights, defined as:
    \[
    z = \mathbf{w}^T \mathbf{X} + b = w_1 x_1 + w_2 x_2 + \dots + w_n x_n + b
    \]
    where:
    \begin{itemize}
        \item \( \mathbf{X} = [x_1, x_2, \dots, x_n] \) are the input features.
        \item \( \mathbf{w} = [w_1, w_2, \dots, w_n] \) are the weights (parameters) associated with each feature.
        \item \( b \) is the bias (intercept term).
    \end{itemize}
    \item \( \sigma(z) \) is the \textit{sigmoid function}, which squashes the output into the range \( (0, 1) \):
    \[
    \sigma(z) = \frac{1}{1 + e^{-z}}
    \]
\end{itemize}
This function creates the S-shaped curve that determines the binary classification. Figure \ref{fig: sigmoidfunction} shows the sigmoid function. 
As a value approaches closer to the 0 or the 1, the probability of a certain classification increases. This model was tested in this study because of its application in sentiment analysis, and also has high success in facial expressions recognition, as shown by Goyani et al \cite{saidi2021}. This may mean that the model will be useful in detecting facial micro-expressions, which was previously mentioned as being vital to detecting deception.

\subsubsection{Random Forrest Classifier}
The second model is the Random Forrest Classifier (RF). In brief, Random Forests are ensemble models that chain together multiple decision trees during training to then merge results, improving accuracy and reducing overfitting \cite{breiman2001}. The decision trees are generated using a bagging algorithm (voting majority). The many decision trees that can go into RF predict an output by forming a “forest” of classifiers that vote for the classification of an input. The RF classifier was considered for this study particularly for the applications of decision trees in sentiment analysis, specifically for speech emotion recognition\cite{saidi2021}. Considering that speech plays a considerable part in lie detection, the RF classifier has potential to increase deception detection accuracy.

\subsubsection{Graph convolutional Network (GCN)}

\begin{figure}[h!]
\centering
\includegraphics[width=0.5\textwidth]{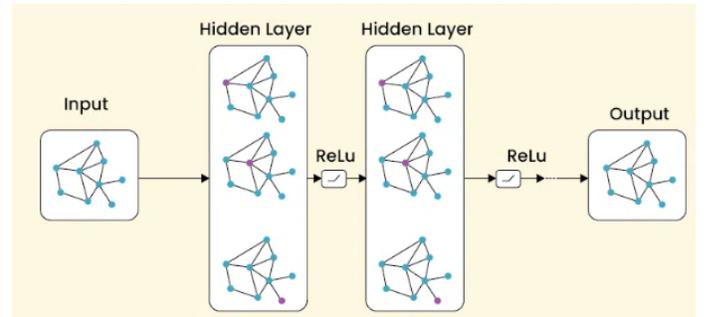}
\caption{shows a diagram for the GCN \cite {GeeksGCN2024}}
\label{fig: diagram for the GCN}
\end{figure}

A graph consists of nodes (vertices) and edges (connections between nodes). In a GCN, each node represents an entity, and the edges represent the relationships between these entities. The primary goal of GCNs is to learn node embeddings, which are vector representations of nodes that capture the graph’s structural and feature information. Figure \ref{fig: diagram for the GCN} shows a diagram for the GCN.

The graphs capture the structural relations among data, harvesting more insights than analyzing data in isolation. However, it is often very challenging to solve the learning problems on graphs, because (1) many types of data are not originally structured as graphs, and (2) for graph-structured data, the underlying connectivity patterns are often complex and diverse \cite {Zhang2019}.

\begin{figure}[h!]
\centering
\includegraphics[width=0.4\textwidth]{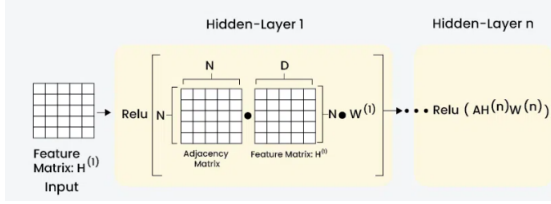}
\caption{shows a basic architecture for the GCN \cite {GeeksGCN2024}}
\label{fig: basic architecture for the GCN}
\end{figure}

A typical GCN architecture contains the following: input layer (initializes the node features); hidden layers (perform the graph convolution operations, progressively aggregating and transforming node features.); output layer (produces the final node embeddings or predictions); fully connected layer (are used at the end of the network to perform tasks such as classification) \cite {GeeksGCN2024}. Figure \ref{fig: basic architecture for the GCN} illustrates the architecture of a GCN.

Spectral-based GCNs leverage the graph Laplacian's eigenvalues and eigenvectors for convolution operations, providing a strong theoretical foundation and capturing global graph structure. The graph Laplacian's eigenvalues represent the frequencies of graph signals, whereas the eigenvectors form a basis for representing functions over the graph, allowing for smooth and meaningful convolutions across the entire graph. This spectral approach effectively captures the global structural information of the graph but can be computationally intensive. Spatial-based GCNs, on the other hand, perform convolutions directly on the graph's local neighborhoods, offering greater flexibility and scalability, making them more suitable for handling large graphs and integrating with other data types \cite {Zhang2019}.

\subsubsection{CNN conv1d}
\begin{quoting}
"Generally, 1D-CNNs are designed to handle one-dimensional data, such as time-series data, sequences (e.g., text), or any data where the primary structure is along a single axis. The kernel (or filter) in a 1D-CNN moves along one dimension. If the data is represented as a vector $[x_1, x_2, \dots, x_n]$, the kernel will slide over this vector to detect patterns within the sequence. The shape of the kernel is a 1D array with dimension $(k,)$, where $k$ is the size of the kernel."
\end{quoting}

\noindent\hfill -- \textit{A. O. Ige and M. Sibiya}, "State-of-the-art in 1D Convolutional Neural Networks: A Survey," \cite{Ige2024}

In a 1D Convolutional Neural Network, the kernel moves along a single axis of the input vector, processing the data effectively. The receptive field of a 1D-CNN kernel involves a contiguous segment of the 1-D input. As the kernel slides across the input, it aggregates information from k consecutive elements. For a given input sequence x and a kernel w, the convolution operation in a 1D-CNN layer can be expressed as \cite{Ige2024}:
\[
(x * w)(t) = \sum_{i=0}^{k-1} x(t + i) \cdot w(i)
\]

where:
\begin{itemize}
    \item $x$ is the 1-d input,
    \item $w$ is the kernel (or filter),
    \item $(x * w)(t)$ denotes the convolution of $x$ and $w$ at position $t$,
    \item $k$ is the size of the kernel,
    \item $x(t + i)$ is the element of the input sequence at position $t + i$,
    \item $w(i)$ is the element of the kernel at position $i$.
\end{itemize}

\begin{figure}[h!]
\centering
\includegraphics[width=0.5\textwidth]{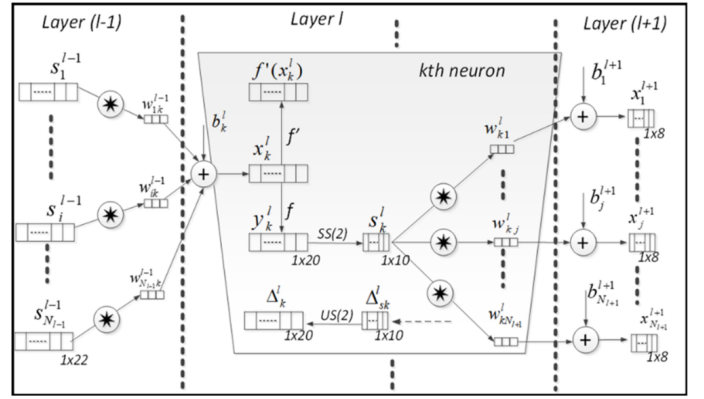}
\caption{Illustration of three consecutive layers in 1D-CNN
 \cite{Ige2024}}
\label{fig: Illustration of three consecutive layers in 1D-CNN}
\end{figure}

An illustration of three consecutive convolutional layers is presented in Fig. \ref{fig: Illustration of three consecutive layers in 1D-CNN}, as seen in \cite{Kiranyaz2015}, where $x_i^k$ is used to denote the input, $b_i^k$ is the bias of the neuron at $k$th position of layer $l$, and the output of the $i$th neuron in the incoming layer $l - 1$ is given as $s_i^{(l-1)}$, and $w_{ik}^{(l-1)}$ denotes the kernel assigned from the neuron in the $i$th position of the first convolutional layer $l - 1$ to the $k$th neuron in the second layer $l$. $y_k^l$ is the intermediate output, $SS$ is the scalar factor used in down sampling, and $f$ is the activation function \cite{Ige2024}.

Forward propagation in 1D-CNN involves passing the input through one or more convolutional layers, pooling layers, and fully connected layers, such that feature map $Z_c$ is given as \cite{Ige2024}:
\[
Z_c = f_c(X * W_c + b_c)
\]
where $X$ is the input, $W_c$ denotes the filter weights, $b_c$ is the bias term, and $f_c$ is the activation function of the convolution, and $X * W_c$ is the convolutional operation between filter weights and bias terms. The spatial dimension of $Z_c$ is reduced by aggregating information from nearby values
through pooling, which is given as \cite{Ige2024}:

\[
A_p = P(Z_c)
\]
where \( P \) is the pooling operation. Then, a fully connected layer \( Z_f \) combines the features learned from the convolutional and pooling operations, and the final activation function \( Y \) is used to obtain the output of the network. Also, backward propagation in 1D-CNN involves computing the gradients of the loss function with respect to the network's parameters, which are used to update the weights and biases. The backward propagation in the fully connected layer is as follows \cite{Ige2024}:

\begin{equation}
    \frac{\partial L}{\partial Z_f} = \frac{\partial L}{\partial Y} \cdot f'_f(Z_f)
    \label{eq:fully_connected_layer}
\end{equation}

Here, \( L \) represents the loss function, and \( f'_f \) denotes the activation function in the fully connected layer. The gradient of the loss with respect to the fully connected weights, \( \frac{\partial L}{\partial W_f} \), is given by \cite{Ige2024}:

\begin{equation}
    \frac{\partial L}{\partial W_f} = \frac{1}{m} \frac{\partial L}{\partial Z_f} \cdot A_p^T
    \label{eq:fully_connected_weights}
\end{equation}

Similarly, the gradient of the loss with respect to the fully connected biases, \( \frac{\partial L}{\partial b_f} \), is calculated as:

\begin{equation}
    \frac{\partial L}{\partial b_f} = \frac{1}{m} \sum \left( \frac{\partial L}{\partial Z_f} \right)
    \label{eq:fully_connected_bias}
\end{equation}

Backpropagation through the pooling layer is performed as outlined in Equation (\ref{eq:fully_connected_bias}) (which is not shown in the image). For the convolutional layer, the backpropagation is given by \cite{Ige2024}:

\begin{equation}
    \frac{\partial L}{\partial Z_c} = \frac{\partial L}{\partial A_p} \cdot P'(Z_c)
    \label{eq:conv_layer_zc}
\end{equation}

The gradient of the loss with respect to the convolutional weights, \( \frac{\partial L}{\partial W_c} \), is expressed as:

\begin{equation}
    \frac{\partial L}{\partial W_c} = \frac{1}{m} X * \frac{\partial L}{\partial Z_c}
    \label{eq:conv_layer_weights}
\end{equation}

Finally, the gradient of the loss with respect to the convolutional biases, \( \frac{\partial L}{\partial b_c} \), is computed as:

\begin{equation}
    \frac{\partial L}{\partial b_c} = \frac{1}{m} \sum \left( \frac{\partial L}{\partial Z_c} \right)
    \label{eq:conv_layer_bias}
\end{equation}

In these equations, \( m \) denotes the batch size, and \( P'(Z_c) \) represents the gradient of the pooling operation. The terms \( \frac{\partial L}{\partial W_c} \) and \( \frac{\partial L}{\partial b_c} \) correspond to the gradients of the loss with respect to the convolutional weights and biases, respectively. These gradients are then used to update the convolutional weights \( W_c \) and biases \( b_c \) using gradient descent \cite{Ige2024}.

\section{Methods}

\subsection{Dataset Collection}
The experiment was conducted under the following guiding question: \textit{Can a multimodal model of facial microexpressions and speech be used to accurately classify the deception of humans?} To successfully research a multimodal model that encompasses both a visual and speech encoder, a dataset containing both video and audio had to be found.

The \textit{Multimodal Real Life Trial Dataset} was used in our experiments, which includes videos and handwritten elements, ideal for in-depth analysis. Each clip was labeled as being deceptive or truthful and had visibility of the face of the speaker as well as the statements spoken by them during the duration of the clip as seen in frames in Fig. \ref{fig: frames from data} \cite{perez2017}. The dataset was composed of 121 testimonies, both truthful and deceptive, that were also manually transcribed and annotated with facial reactions.

\begin{figure}[h!]
\centering
\includegraphics[width=0.4\textwidth]{frames from data.png}
\caption{The dataset sample frames, pulled from [6]’s dataset, display hand movements, microfacial expressions, and facial reactions.
}
\label{fig: frames from data}
\end{figure}

The videos in this dataset had an average length of 28.0 seconds, with the deceptive videos averaging 27.7 seconds and the truthful videos averaging 28.3 seconds. The 56 distinct speakers in these clips were made up of 21 female and 35 male speakers, each between the ages of 16 to 60 \cite{perez2017}. This dataset was found on the University of Michigan’s Deception Detection and Misinformation datasets list. Fig. \ref{fig: recorde annotation distribution} shows the distribution of the recorded annotations in the dataset.

\begin{figure}[h!]
\centering
\includegraphics[width=0.5\textwidth]{recorde annotation distribution.png}
\caption{The dataset sample frames, pulled from [6]’s dataset, display hand movements, microfacial expressions, and facial reactions.
}
\label{fig: recorde annotation distribution}
\end{figure}

\subsection{Data Preprocessing (OpenSmile, ViT, manual annotations)}
In this case, using audio, video, and textual analysis to predict the result can give us a more accurate picture than rudimentary polygraphs and much more freedom to innovate over related works and previous model pipelines. Primarily, to be able to fully grasp the importance of audio and video from the datasets, visual and auditory patterns can be pulled out using two models: Vision Transformer (ViT) and OpenSmile.

ViT is an image recognition encoder. It was used to extract features from the visual data. ViT split the individual image frames of the video at a sampling rate of 50 Hz into different patches. These patches were both linearly embedded with patch and position embeddings. ViT’s benefits of computational efficiency and accuracy come to light when iterating over large datasets, and thus, we use a pretrained ViT to fit this task so that it can outperform models like ResNet and other CNNs as mentioned in \cite{dosovitskiy2021}. The vectors from this linear projection were then interpolated to match the dimensions of audio and text features, which then will be saved for concatenation \cite{dosovitskiy2021}.

Further, to satisfy the multimodal label, OpenSmile is our chosen processor, working in tandem with models like a CNN and simpler models mentioned below. OpenSmile, or Open-source Speech and Music Interpretation by Large-space Extraction, was first developed at the Technical University of Munich to attempt to create SEMAINE, a fully socially conscious software \cite{schroder2010}. OpenSmile’s main function for that software is for audio and emotional analysis and feature extraction, which in this paper’s use case works well for identifying abnormalities in tone, hesitation, and emotional nuance between inputs of lie and truth \cite{eyben2010}. Via the clips provided in the dataset, the “ffmpeg” package is used to successfully extract audio from the videos in a “.wav” format at a sampling rate of 50 Hz, which will aid in concatenation with other features. In the context of this research, OpenSmile takes in the wave-form input audio files and analyses features like pitch, loudness energy, and MFCC (Mel-frequency cepstral coefficients). Using the aforementioned features, aspects like tone, rhythm, and timbre will be saved in vectors to help classify our data into buckets of truths and lies \cite{eyben2010}.

Finally, the features saved into vector format from OpenSmile, ViT, and the handwritten annotations can be combined using simple concatenation. Then those are converted into Tensor format for experimentation with various models.

\subsection{Dataset Curation and Analysis}
The dataset was further curated to fit the requirements of the project. Since there was an imbalance in the number of truthful and deceptive videos, one of the deceptive videos was dropped from the dataset at random to create an even ½ split of 60 truthful to 60 deceptive videos. The dataset was then processed by splitting audio files from the trial videos using FFmpeg, followed by OpenSmile to extract features from the audio. ViT was used to extract individual frames from the videos and extract features from them.

The file names of the different extracted feature files were placed in different split files for training, testing, and validation. The train set contained 70\% of the dataset, the validation set contained 10\% of the dataset, and the test set contained 20\% of the dataset. These file names were used to identify which extracted feature files to pull when training, validating, and testing.

\begin{figure}[h!]
\centering
\includegraphics[width=0.5\textwidth]{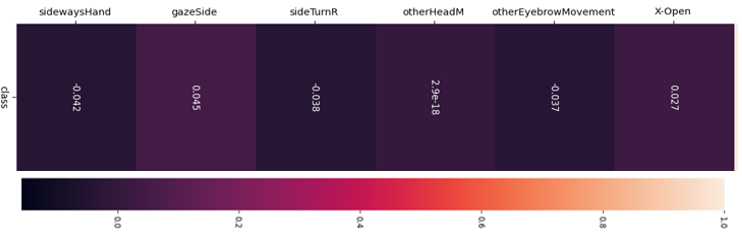}
\caption{shows the correlation heatmap for some sample columns.
}
\label{fig: heatmap}
\end{figure}

\begin{figure}[h!]
\centering
\includegraphics[width=0.4\textwidth]{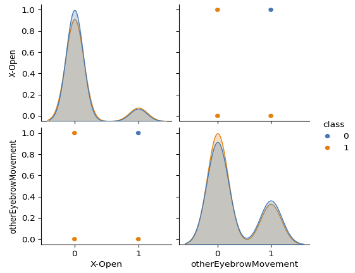}
\caption{shows the KDE for some columns
}
\label{fig: kde}
\end{figure}

For the CSV, the data was analyzed using pandas and seaborn. A heatmap was done to determine the features that don’t have a strong correlation with the target, as shown in Fig. \ref{fig: heatmap}. Also, KDE using seaborn was done whereas making \texttt{hue = ‘class’} to see the data distribution as illustrated in Fig \ref{fig: kde}. According to that analysis, any column that correlated less than 0.05, as shown in columns samples in Fig. \ref{fig: heatmap}, and the distribution of the classes (deception and truthful (0,1)) is approximately the same, as shown in the columns samples in Fig. \ref{fig: kde}, has been dropped. ( Note: The analysis (heatmap and KDE) was done to all columns, but here some samples were mentioned instead of all since if all columns were mentioned in the heatmap or KDE, that would take a big place. The full heatmap and KDE are in the file named "data analysis" that was uploaded on GitHub; you can access the link in the appendix section)

\subsection{Models}

\subsubsection{Convolutional Neural Networks}
Firstly, a Convolutional Neural Long Short-Term Memory Network (CNN LSTM), is a model used in our experimentation. CNN LSTM models specialize in focusing on spatial relations in images, describing videos, and actions with text, and classification \cite{donahue2014}. In this case, the CNN LSTM takes in concatenated data of audio, visual, and annotated features and then classifies that data as a truth or a lie.

The second iteration of CNNs is a basic, custom Convolutional Neural Network. Using a high variety of layers within a 1-dimensional CNN framework, the model can match well with the results of data preprocessing. Further, to regulate and standardize the data to avoid overfitting because of the lack of samples, a Dropout function and Max Pooling function, as seen in Equation \ref{eq:maxpool}, are used when needed in the layers.

Further, for the classification aspect of the complex CNN, a classical Sigmoidal function is used from the neural network python module since Sigmoid is good in binary classification; similarly, the loss function was Binary Cross Entropy. Equation \ref{eq:sigmoid} shows the Sigmoid function, whereas Equation \ref{eq:BCE} shows the Binary Cross Entropy loss function.

\begin{equation}
\mathbf{F}_{\text{max}(\mathbf{x})} = \max \left\{ x_i \right\}_{i=0}^{N}
\label{eq:maxpool}
\end{equation}

\begin{equation}
\text{Sigmoid}(x) = \sigma(x) = \frac{1}{1 + \exp(-x)}
\label{eq:sigmoid}
\end{equation}

\begin{equation}
\text{BCE} = -\frac{1}{N} \sum_{i=1}^{N} \left[ y_i \log(p_i) + (1-y_i) \log(1-p_i) \right]
\label{eq:BCE}
\end{equation}

\subsubsection{Classical Machine Learning Models}
Furthermore, based on the characteristics of the dataset, pre-processing output tensors, and the number of samples, more simplistic classification models can be used in this pipeline to gain insight into their effectiveness with lie detection. A Random Forest Classifier was used initially, in tandem with previous papers such as \cite{soldner2019} mentioned previously. 

Meanwhereas, Logistic Regression is a statistical binary classification tool that applies a logistical function to a linear arrangement of input features. Doing so helps serve as a baseline for further models and is easily interpretable.

\subsubsection{Experimental FiLM and Graph Convolutional Networks}
The first experimental method integrates audio and visual data using speech and vision encoders combined with Feature-wise Linear Modulation (FiLM) \cite{perez2017}. In this approach, audio signals are processed using a speech encoder that combines CNNs and Transformers to extract a sequence of hidden vectors to begin classification \cite{reddy2019}. This fusion approach enhances the system's ability to capture the nuanced interactions between audio and visual cues by gaining the capability to use visual reasoning and classification, which are critical for accurate lie detection \cite{sun2021}.

The second method employs Graph Convolutional Networks (GCNs) focusing on scattered (spatially apart) features in the image by creating graphs from the features supplied. The processed audio features are integrated with the GCN and Transformer outputs using a suitable fusion method. This combination leverages the strengths of GCNs in handling complex graph structures and Transformers in modeling long-range dependencies, making it highly effective for detailed facial movement analysis \cite{plosone2021}. In this instance, a specific iteration, spectral-based GCNs were used. First, Spectral-based GCNs take advantage of the graph Laplacian's eigenvalues and eigenvectors to define convolutions in the spectral domain, capturing global graph structures and relationships. Second, their usage for spectral graph theory provides a robust theoretical foundation for processing the graph data, helping optimize the model for detecting subtle patterns in multimodal data. Third, since the multimodal model integrates features from different sources (e.g., auditory and visual), spectral-based GCNs can facilitate feature integration by providing a global view of the graph structure. Although the model training may be high computationally cost, the model was trained on a server to overcome this challenge, increasing the thorough nature of this approach \cite {Zhang2019}. This specific adaptation of Graphic Convolutional Networks is fully experimental.

\section{Results}

\begin{table}[h!]
\centering
\caption{Classification reports for different models}
\label{table: classification reports table}

\begin{tabular}{|c|c|c|c|c|}
\hline
\textbf{Model}           & \textbf{Class}    & \textbf{Precision} & \textbf{Recall} & \textbf{F1-score} \\ \hline
\multirow{2}{*}{Random Forest} & Deception & 0.77      & 0.91   & 0.83    \\ \cline{2-5} 
                      & Truthful  & 0.83      & 0.73   & 0.8     \\ \hline
\multirow{2}{*}{Logistic Regression} & Deception & 0.77      & 0.91   & 0.83    \\ \cline{2-5} 
                      & Truthful  & 0.89      & 0.73   & 0.8     \\ \hline
\multirow{2}{*}{GCN}   & Deception & 1         & 0.07   & 0.14    \\ \cline{2-5} 
                      & Truthful  & 0.48      & 1      & 0.65    \\ \hline
\multirow{2}{*}{CNN conv1d} & Deception & 0.91      & 1      & 0.95    \\ \cline{2-5} 
                      & Truthful  & 1         & 0.9    & 0.95    \\ \hline
\end{tabular}
\end{table}

\begin{table}[h!]
\centering
\caption{Test accuracy for each model after each fold}
\label{table: Accuracy after fold}

\resizebox{0.5\textwidth}{!}{
\begin{tabular}{|c|c|c|c|c|c|c|c|}
\hline
\textbf{Model} & \textbf{1\textsuperscript{st} fold} & \textbf{2\textsuperscript{nd} fold} & \textbf{3\textsuperscript{rd} fold} & \textbf{4\textsuperscript{th} fold} & \textbf{5\textsuperscript{th} fold} & \textbf{Mean} & \textbf{Std} \\ \hline
\textbf{Random forest} & 0.59 & 0.8636 & 0.6818 & 0.619 & 0.8095 & 0.712 & 0.107 \\ \hline
\textbf{Logistic regression} & 0.772 & 0.7727 & 0.7727 & 0.7142 & 0.7142 & 0.7402 & 0.0269 \\ \hline
\textbf{CNN conv1d} & 0.909 & 0.909 & 1 & 0.95 & 1 & 0.954 & 0.04 \\ \hline
\end{tabular}
}

\end{table}

\begin{figure}[h!]
\centering
\includegraphics[width=0.5\textwidth]{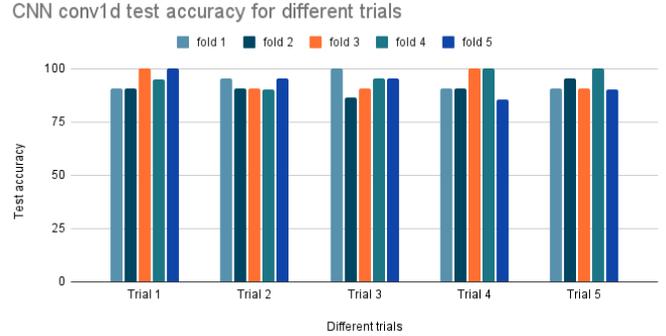}
\caption{Shows different trials for the best model (conv1d)}
\label{fig: CNN conv1d test accuracy for different trials}
\end{figure}

\begin{figure}[h!]
\centering
\includegraphics[width=0.5\textwidth]{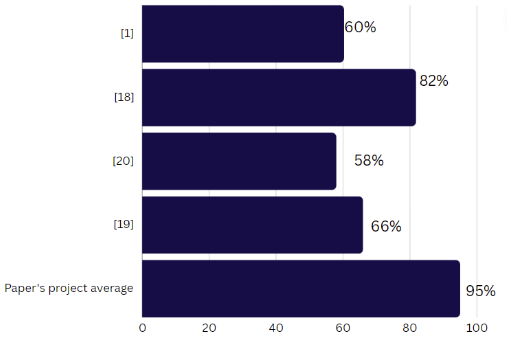}
\caption{shows different accuracy of previous papers}
\label{fig:comparison}
\end{figure}

As discussed in the methods, this paper examined multiple models and K-folds. Table \ref{table: Accuracy after fold} shows the used models with the accuracy of each fold and the mean and standard deviation of all folds of each model. Table \ref{table: classification reports table} illustrates the classification report of each model. Figure \ref{fig: CNN conv1d test accuracy for different trials} shows different trials for the best model (conv1d).

Figures \ref{fig:comparison} and \ref{fig: CNN conv1d test accuracy for different trials}, along with tables \ref{table: classification reports table}, \ref{table: Accuracy after fold} and \ref{tab:lie_detection_accuracy}, answer the first research question of the paper that is "How
effective is the proposed AI model in detecting lies compared
to traditional methods and some recent AI models?"

Figures \ref{fig:deceptive sample}, \ref{fig:truthful sample}, and \ref{fig:shap} answer the second research question of the paper "Which features carry the highest weights in prediction?"

\section{Discussion}
\subsection{Interpretation of the collected results}

The following are the commonly used evaluation metrics in classification reports: Precision is the ratio of correctly predicted positive observations to the total predicted positives. The metric helps see the true pinpointed accuracy of positive results whereas gauging false positives as well. Second, recall is the ratio of correctly predicted positive observations to all observations in the actual class. In the context of the task, it answers the question: "Of all the instances that were lies, how many were correctly predicted as a lie?" Finally, the F1-score is the harmonic mean of Precision and Recall. It provides a single metric that balances both concerns, especially when you have an imbalanced dataset where one class is more prevalent. 

The illustrated results of the conv1d model in Tables \ref{table: classification reports table} \ref{table: Accuracy after fold}, which demonstrate the classification report and accuracy’s mean and standard deviation, and the illustrated performance in each trial in Fig. \ref{fig: CNN conv1d test accuracy for different trials}, where each trial has 5 folds, ensure the model's reliability and accuracy. For example, the classification report of the conv1d shows no bias in the model, the standard deviation is 0.04, which is relatively low, and the mean is 95.4\%, which is a respected accuracy in the context of lie detection using AI. In contrast, other, more typical models like Logistic Regression or Random Forest Classifiers faced a lower average precision. The F-1 scores of both were close to equal, with the lie class at 83\% and 80\% for the truth class. Further, the spectral GCN model did not prove as successful, as seen by the difference in precision, one class reached 100\% whereas the other was significantly lower, hinting at an extreme bias towards a lie or a truth. Yet, when we tried the CNN conv1d on the training set without manual annotation (only audio and visuals were used), the model’s average accuracy was still 95+\%.

\subsection{The proposed model vs. previous studies}
Fig. \ref{fig:comparison} shows where our model stands in comparison to previous studies. The figure contains the accuracy of different studies, considering that they may use the same data or different ones. These figures help to answer the first research question of this paper: “How effective is the proposed AI model in detecting lies compared to traditional methods and some recent AI models?” For example, although the study \cite{dhabarde2023hybrid} used a fusion of two different AI models, CNN and support vector machine, increasing the complexity and the need for high computational resources, our solution’s best model used only conv1d and achieved higher accuracy. Its simplicity, along with higher accuracy proves it to be a more effective model overall, especially in real-world situations.

\subsection{Interpretation of the proposed model}
\begin{figure}[h!]
\centering
\includegraphics[width=0.5\textwidth]{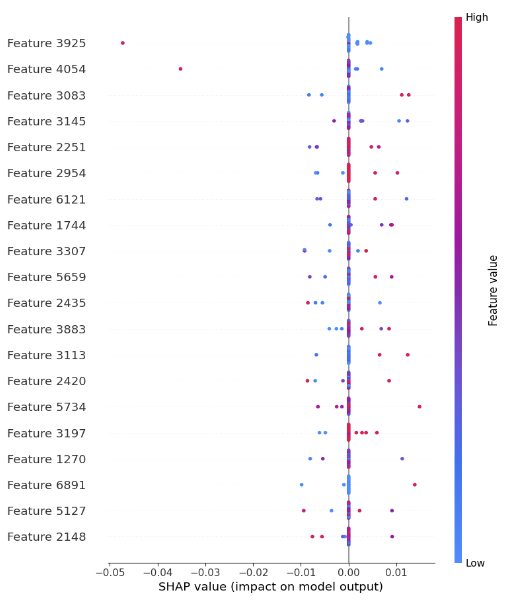}
\caption{illustrate the most and least significant features regarding the paper's model}
\label{fig:shap}
\end{figure}

Explainable AI (XAI) was used to interpret our model. For example, to explain the predictions of a machine learning model, SHAP (SHapley Additive exPlanations), a popular method for interpreting complex models, was used. It was used to generate a summary plot of the SHAP values, which visually represents the impact of each feature on the model's predictions. The plot helps in understanding which features are most influential and how they contribute to the model's output. The output is in Fig. \ref{fig:shap}, illustrating that Features 3925 and 4054 are the most important. They are audio features. (Note: since this requires a lot of computation, we used a subset (which contains both target classes) of the data for visualization.)

\begin{figure}[h!]
\centering
\includegraphics[width=0.5\textwidth]{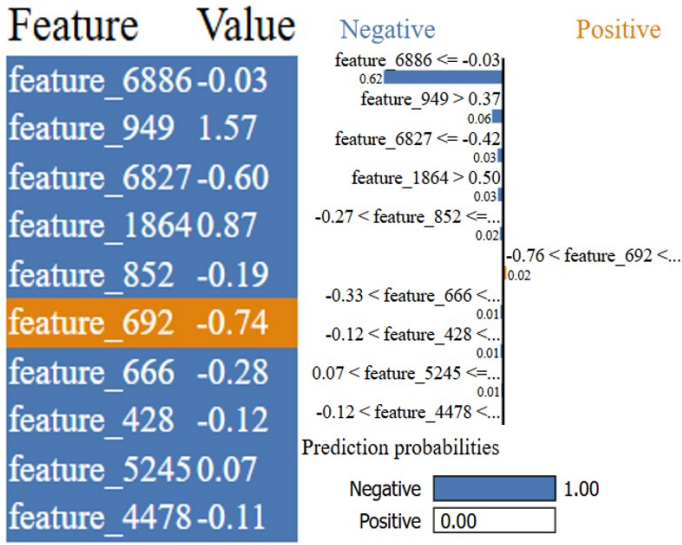}
\caption{LIME's output on a deceptive sample}
\label{fig:deceptive sample}
\end{figure}

\begin{figure}[h!]
\centering
\includegraphics[width=0.5\textwidth]{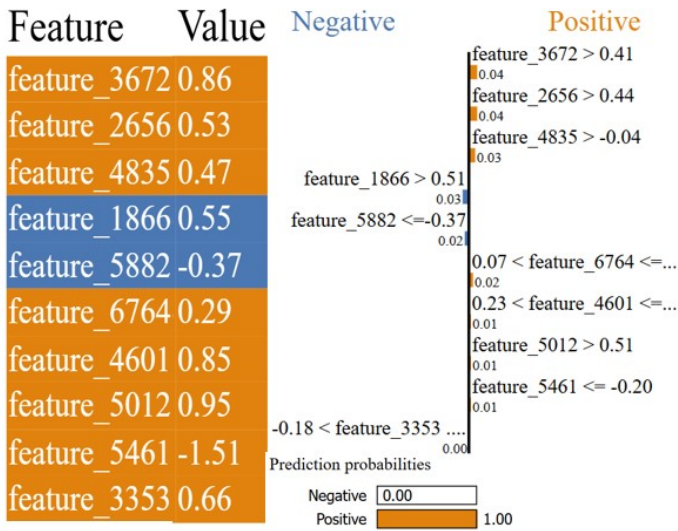}
\caption{LIME's output on a truthful sample}
\label{fig:truthful sample}
\end{figure}

The LIME (Local Interpretable Model-agnostic Explanations), an XAI technique, framework explains the predictions of a machine learning model. LIME provides insight into why a model made a specific prediction by approximating the model locally with an interpretable model. It provides a local interpretation of a specific prediction made by a neural network model. By focusing on one instance and showing how different features contributed to the prediction, LIME helps to make complex models more understandable, especially in a classification context. We used two samples, one for deception (see Fig. \ref{fig:deceptive sample}) and one for truthful (see Fig. \ref{fig:truthful sample}).

By analyzing Fig. \ref{fig:deceptive sample}, feature 949 had the highest positive contribution (1.57) towards the "Negative" (truthful) class, indicating that when this feature is at a higher value, the model is more confident that the instance is not associated with lying. On the other hand, feature 692 and feature 6827 had significant negative contributions (-0.74 and -0.60, respectively), suggesting that when these features have lower values, the model is more likely to classify the instance as "Negative" (truthful). The model is highly confident in predicting the "Negative" class (with a probability of 1.0), which is visualized by the zero probability for the "Positive" class. This suggests a clear decision made by the model based on the combination of feature values.

From Fig. \ref{fig:truthful sample}, 3672, 2656, and 4835 features had values that strongly contributed to the "Positive" prediction, indicating they were key indicators of deception according to the model. Features 1866 and 5882 contributed towards a "Negative" prediction (indicating truth) but were outweighed by the features supporting a "Positive" prediction.

The mentioned features 3925, 4054, 949, 3672, 2656, 4835, 1866, and 5882 are related to audio. On the other hand, feature 6827 is related to the visuals that were extracted from videos. CNN conv1d specializes in audio-related tasks, thus, it relays more audio features than visual features.

\subsection{Limitations and Recommendations}
There were certain limitations of this study that affected the results of this experiment. Primarily, the lack of a high-quality and quantity dataset with both video and audio data was profound. The shortage of data in the Real-life Trial dataset may have decreased the overall accuracy of the model on the dataset. The dataset only contained 121 videos, of which 120 were used. 5 K-folds were used to split up the training and validation to avoid overfitting, but the small dataset still proved difficult when trying to train data. Furthermore, as Mambreyan et al. (2022) have shown, the Real-life Deception Dataset has significant gender bias that classifiers may exploit. Other larger datasets usually have features that are only manually annotated or are already extracted for features. Expanding sampling to other scenarios, and diversifying subjects based on gender, ethnicity, and beliefs or ideologies could be vital in a more universal model. Finding these higher-quality datasets with more data samples would increase the amount of training data available, reduce social bias, and thereby increase the accuracy of the experiment.

That said bias appeared in the tests of the GCN model as stated above. Based on the data in Table \ref{table: classification reports table}, it is evident that the GCN model shows a preference for class 1, shown by its recall but low precision for class 1 and notably low recall for class 0, indicating extreme class bias. The spectral-based GCN showed much promise in the previous experiments above but fell short, yet we encourage it to be experimented on further. To enhance the model's accuracy, especially when utilizing the spectral GCN model, it will be beneficial to present more data, employ methods of resampling, and provide higher computing power for those tasks.

\section{Conclusion}

The results of this study highlight the critical role of using multimodal feature extraction techniques for advancing lie detection technologies. By leveraging audio features via OpenSmile, visual data through Vision Transformer, and transcriptions of gestures and micro-expressions, the CNN Conv1D model achieved a high accuracy of 95.4 \%, surpassing many state-of-the-art approaches. Even without manual transcriptions, the model performed admirably at 95 \%, demonstrating the robustness of the architecture, particularly with a limited dataset. Both audio features and visual features were vital to the performance of CNN Conv1D, and its success. Furthermore, the paper addressed the recommendation, which focuses on incorporating multi-modal data, mentioned by a study’s \cite{Dhabarde} author.
Despite these promising results, this study's limitations—particularly the small and homogeneous dataset—underscore the necessity for further research. It is crucial to expand dataset diversity to include participants from various demographic groups and scenarios. Explainable AI (XAI) revealed that audio and visual features were the most significant contributors to the model's decisions, indicating the importance of focusing on these modalities in future studies. However, integrating additional modalities like thermal imaging, heart rate monitoring, and moisture tracking could further enhance model performance, especially in complex real-world applications.
Addressing these challenges will not only improve model generalizability but also help mitigate ethical concerns related to biases in facial micro-expression recognition across different racial and gender groups. Future work should explore ablation studies and alternative architectures to deepen our understanding of how multimodal learning can be optimized. By continuing to build on this research, we move closer to creating an accurate, reliable, and ethical alternative to traditional polygraph tests, with potential applications in criminal justice and law enforcement.

\bibliographystyle{plain}

\appendix
\section{Appendix}

You can find the code for the Multi-modal Lie Detection Project at the following GitHub repository:

\href{https://github.com/AbdelrahmanAbdelwahab1/Multi-modal-Lie-detection-project}{https://github.com/AbdelrahmanAbdelwahab1/Multi-modal-Lie-detection-project}

\end{document}